\documentclass[12pt]{article}
\input epsf.sty
\usepackage{mathrsfs}
\usepackage{amsmath}
\usepackage{mathrsfs}
\usepackage{amssymb}
\usepackage{color}
\usepackage{psfrag}
\usepackage{graphicx}
\usepackage{subfigure}
\usepackage{rotating}

\newcommand{\bea}{\begin{eqnarray}}
\newcommand{\eea}{\end{eqnarray}}
\newcommand{\be}{\begin{equation}}
\newcommand{\ee}{\end{equation}}
\newcommand{\vs}[1]{\vspace{#1 mm}}

\newcommand{\dsl}{\pa \kern-0.5em /}

\newcommand{\half}{\frac{1}{2}}
\newcommand{\pa}{\partial}

\newcommand{\nn}{\nonumber\\}

\begin{document}
\topmargin 0pt
\oddsidemargin 0mm

\begin{flushright}

%USTC-ICTS-10-12\\

%hep-th/yymmnnn\\

\end{flushright}

\vspace{2mm}

\begin{center}
{\Large \bf Holographic entanglement entropy\\ of the near horizon 1/4 BPS
F-D$p$ bound states}

\vs{10}

{Parijat Dey\footnote{E-mail: parijat.dey@saha.ac.in} and
Shibaji Roy\footnote{E-mail: shibaji.roy@saha.ac.in}}

 \vspace{4mm}

{\em

 Saha Institute of Nuclear Physics,
 1/AF Bidhannagar, Calcutta-700 064, India\\}

\end{center}

\vs{10}

\begin{abstract}
It was shown in \cite{Dey:2012tg} that the near horizon limit of
the 1/4 BPS threshold F-D$p$ (for $0\leq p \leq 5$, $p \neq 4$) bound state 
solutions of type II string
theories give rise to space-time metrics endowed with Lifshitz
scaling along with hyperscaling violation. Here we compute the
holographic entanglement entropy of this system for all $p \neq 4$
(for $p=4$ the space-time has AdS$_2$ structure). For $p=3,5$,
we get the expected area law behavior of the entanglement entropy.
For $p=0,1$, the entanglement entropy has new area law violations 
and has the behavior which is in between the linear and logarithmic behaviors. 
For $p=2$, we get a logarithmic violation of the area law. We also
compute the entanglement entropy at finite temperature and show that
as the temperature rises, the entanglement entropy makes a crossover
to the thermal entropy of the system. We thus obtain the string theoretic realization
of holographic EE and various of its aspects noted earlier for generic metric
with hyperscaling violation. 
\end{abstract}

\newpage

\section{Introduction}

An entanglement entropy (EE) is inherently a quantum mechanical concept
and is defined as the von Neumann entropy of the reduced density 
matrix of a subsystem $A$ of the full quantum system obtained by taking 
the trace over the degrees of freedom of the subsystem $B$ complement to $A$.
This measures how the two subsystems $A$ and $B$ are entangled with 
each other (see, for example, \cite{Bombelli:1986rw, Srednicki:1993im,
Holzhey:1994we, Calabrese:2004eu, Eisert:2008ur, Nishioka:2009un,
Takayanagi:2012kg}, including some reviews).
The concept has found many applications in various branches of physics,
particularly, in condensed matter physics \cite{Vidal, Peschel, Its,
  Kitaev:2005dm, Levin}. It has been realized in the
study of condensed matter systems that to describe quantum phases of
matter and their transitions which occur at nearly zero temperature, 
the relevant quantity to characterize these phases are their patterns of
entanglement and not the conventional order parameters \cite{Vidal,
  Kitaev:2005dm, Levin}. Therefore, EE
is potentially useful to study systems at or near quantum criticality.

For low dimensional ($d < 2$) quantum field theory or quantum many 
body systems it is known \cite{Holzhey:1994we, Calabrese:2004eu} how 
to compute the EE, however, for higher 
dimensional ($d\geq 2$) quantum theory the computation is not easy for 
a generic 
subsystem $A$ even for free theories. Motivated by this and also by
looking at the similarity with the Bekenstein-Hawking black hole 
entropy, Ryu and Takayanagi \cite{Ryu:2006bv, Ryu:2006ef} gave a 
prescription to obtain the EE in any
dimensions using the holography and AdS/CFT \cite{Maldacena:1997re, 
Aharony:1999ti}. The idea is that 
the EE of a region $A$ (with boundary $\partial A$) in a time slice of 
$(d+1)$ dimensional CFT can be calculated by the minimal area of the
manifold $\gamma_A$ embedded in the time slice of the bulk AdS$_{d+2}$ 
spacetime, such that $\partial \gamma_A = \partial A$, and is given
by the relation \cite{Ryu:2006bv}
\be\label{EE}
S_A = \frac{{\rm Area}(\gamma_A^{\rm min})}{4 G_N^{d+2}}
\ee
In \eqref{EE} $\gamma_A^{\rm min}$ denotes the manifold with minimal area 
whose boundary
is $\partial A$. Also $G_N^{d+2}$ is the $(d+2)$ dimensional Newton's
constant. The expression in \eqref{EE} is known as the holographic
entanglement entropy and in lower dimensions it has been checked to match
with the known results of a quantum system \cite{Ryu:2006bv}. In higher 
dimensions
also $S_A$ calculated from the gravity side can be seen to give correct
qualitative behavior expected from field theory results. 

Although originally the holographic EE was given using AdS/CFT, but it is
believed to hold even for the non-AdS/non-conformal correspondence or in 
general for gauge/gravity duality \cite{Klebanov:2007ws, Pakman:2008ui,
Arean:2008az, Ramallo:2008ew, Asplund:2011cq, vanNiekerk:2011yi}. Holographic 
EE has been shown to give 
the correct behavior even for the gravity theories having Lifshitz scaling
symmetry as well as theories with hyperscaling violation 
\cite{Azeyanagi:2009pr, Shaghoulian:2011aa, Dong:2012se, Kim:2012nb}. As some condensed
matter systems near quantum critical point show such scaling behavior
\cite{Sachdev, Sachdev:2012dq}, these
gravity theories may describe the dual of those condensed matter systems 
and therefore studying their
holographic EE can help us to understand various quantum phases of these
systems. This has been addressed in some recent papers \cite{Dong:2012se,
Ogawa:2011bz, Huijse:2011ef, Kim:2012nb}.

In this paper we compute\footnote{We would like to emphasize that
the holographic EE has been computed for the generic Lifshitz metric in
\cite{Azeyanagi:2009pr} (the special case of D3-D7 scaling solution is also
discussed there) and for a generic metric with hyperscaling violation in
\cite{Dong:2012se}. However, it is not clear in \cite{Dong:2012se} how such
a space-time, particularly, a Lifshitz metric with hyperscaling violation
can be obtained from any fundamental theory. We here use the solutions of the
type F-D$p$, as mentioned below, of string theory and show how the various
properties of the EE observed in \cite{Dong:2012se} can be realized for
different values of $p$.} the holographic EE for the geometries
obtained by taking the near horizon limit of 1/4 BPS, threshold F-D$p$ bound
state solutions of type II string theories. In an earlier paper 
\cite{Dey:2012tg} we have shown
that these geometries show a Lifshitz scaling symmetry along with a
hyperscaling violation and may be dual to some condensed matter systems near
critical point. We find that for $p=3,\,5$, the holographic EE has the usual
area law behavior \cite{Bombelli:1986rw, Srednicki:1993im}. However, 
for $p=0,\,1,\,2$, the area law is violated 
indicating new
phases in dual field theory. For $p=0,\, 1$, we find that the holographic EE
has a behavior which is in between linear and logarithmic behaviors shown
earlier \cite{Dong:2012se} for a system with hyperscaling 
violations. Here
we find a string theory realizations of that. For $p=2$, we get a logarithmic
violation of the area law indicating that the corresponding dual system 
represents compressible metallic states with hidden Fermi surface 
\cite{Ogawa:2011bz, Huijse:2011ef}. We also
compute the holographic EE at finite temperature by extending our results to
the non-extremal F-D$p$ bound state solutions. We study both the low
temperature and the high temperature behavior of this EE and find that at high
temperature the EE makes a crossover to the thermal entropy of the system 
\cite{Swingle:2011mk}.

This paper is organized as follows. In section 2, we give the computation of
holographic EE for the dimensionally reduced, near-horizon F-D$p$ geometry
and discuss their behaviors in various cases. In section 3, we discuss the
finite temperature extension of the holographic EE and discuss the low and
high temperature behaviors. We conclude in section 4.    

\section{Entanglement entropy of F-D$p$ system}

In \cite{Dey:2012tg} we constructed the 1/4 BPS, threshold F-D$p$ 
(for $0 \leq p \leq 5$) bound state
solutions of type II string theories. In the near horizon limit the string
frame metric, the dilaton and the form-fields of these solutions in a 
suitable coordinate take the form 
\bea\label{fdpinnewcoord}
ds^2 &=& Q_2^{\half} u^{\frac{2-p}{4-p}}\left[-\frac{dt^2}{Q_1Q_2
u^{\frac{4(5-p)}{4-p}}}
+ \frac{\sum_{i=1}^p(dx^i)^2}{Q_2 u^2} + \frac{(dx^{p+1})^2}{Q_1 u^2}
+ \frac{4}{(4-p)^2}\frac{du^2}{u^2} + d\Omega_{7-p}^2\right]\nn
e^{2\phi} &=& \frac{Q_2^{\frac{3-p}{2}}}{Q_1} u^{\frac{(6-p)(1-p)}{(4-p)}}\nn
B_{[2]} &=& -\frac{1}{Q_1 u^{\frac{2(6-p)}{4-p}}} dt \wedge dx^{p+1},
\qquad A_{[p+1]}\,\, =\,\, -\frac{1}{Q_2 u^{\frac{2(6-p)}{4-p}}} dt \wedge dx^1
\wedge \cdots \wedge dx^p
\eea
Here $Q_1,\,Q_2$ are the F-string and D$p$-brane charges
respectively. $B_{[2]}$ is the NSNS 2-form which couples to F-string and
$A_{[p+1]}$ is the RR $(p+1)$-form which couples to D$p$-brane. Note that
in \eqref{fdpinnewcoord} $p \neq 4$, since for $p=4$, the metric has AdS$_2$ 
structure. Also in the above while for $p < 4$, $u \to 0$ corresponds to 
going to the UV and $u \to \infty$ corresponds to going to the IR, for $p >4$,
$u \to 0$ corresponds to going to the IR and $u \to \infty$ corresponds to
going to the UV. It is clear from \eqref{fdpinnewcoord} that the part of the
metric in square bracket is invariant under the scale transformations
\be\label{scaletr}
t \to \lambda^{\frac{2(5-p)}{4-p}} t \equiv \lambda^z t, \quad 
x^{1,2,\ldots,(p+1)}\to \lambda x^{1,2,\ldots,(p+1)}, \quad u \to \lambda u
\ee
where $\lambda$ is a scale parameter and $z$ is the dynamical scaling 
exponent and has a value $2(5-p)/(4-p)$. However, the full metric is not 
scale invariant and therefore it has a
hyperscaling violation. Now since the dilaton in this case is not constant, in
order to find the hyperscaling violation exponent we have to compactify
the metric on $S^{7-p}$ and then express the $(p+3)$-dimensional reduced 
metric in Einstein frame. It has the form,   
\be\label{reducedmetric}
ds_{p+3}^2 = Q_1^{\frac{2}{p+1}} Q_2
u^{2\frac{p(4-p)-(p-2)}{(4-p)(p+1)}}\left[-\frac{dt^2}{Q_1Q_2
u^{\frac{4(5-p)}{(4-p)}}} + \frac{\sum_{i=1}^p(dx^i)^2}{Q_2u^2} + 
\frac{(dx^{p+1})^2}{Q_1u^2} + \frac{4}{(4-p)^2}\frac{du^2}{u^2}\right]
\ee
Now it is clear that the metric transforms under \eqref{scaletr} as,
\be\label{hyper}
ds_{p+3} \to \lambda^{\frac{p(4-p)-(p-2)}{(4-p)(p+1)}} ds_{p+3} \equiv
\lambda^{\theta/d} ds_{p+3} 
\ee
where $\theta$ is the hyperscaling violation
exponent which has the value $p-(p-2)/(4-p)$ in this case and $d=p+1$ 
is the spatial dimension of the boundary theory. 

In this section we compute 
the EE on a strip embedded on the boundary of the fixed time slice of the
$(p+3)$-dimensional bulk
geometry given in \eqref{reducedmetric} using the proposal of Ryu and 
Takayanagi \cite{Ryu:2006bv, Ryu:2006ef}. We note that the supergravity 
solution \eqref{fdpinnewcoord}
is valid as long as the effective string coupling $e^{\phi} \ll 1$ and
the curvature of the metric ${\cal R} \ll \ell_s^{-1}$, where $\ell_s$ is the
fundamental string length. This gives restrictions on the radial parameter 
$u$ (which is related to the energy parameter in the boundary theory) in 
terms of the charges of the F-string and D$p$-branes and we will discuss 
them later.

To find the holographic EE, we first calculate the minimal area of the
manifold embedded in the time slice of the background \eqref{reducedmetric}
bounded by the edge of $A$, i.e., $\partial A$ which is a strip given by
$-\ell \leq x^{p+1} \leq \ell$ and $0\leq x^i \leq L$, with 
$i=1,\,2,\,\ldots,\,p$. The area of the embedded manifold is,
\be\label{area}
{\rm Area}(\gamma_A) = \int d^{p+1}x \sqrt{g} 
\ee
where $g$ is the determinant of the metric induced on $\gamma_A$. For the
strip just mentioned, the above area reduces to
\be\label{areareduced}
{\rm Area}(\gamma_A) = (Q_1Q_2)^{\frac{1}{2}}L^p \int_{-\ell}^{\ell} dx^{p+1} 
u^{-\frac{2}{4-p}} \sqrt{1+\frac{4Q_1{\dot u}^2}{(4-p)^2}}
\ee
where $\dot u = \partial u/\partial x^{p+1}$ and $u(x^{p+1})$ gives the 
embedding of the edge of the strip into the time slice of the bulk. Note that
here the boundary is located at $u=0$ for $p<4$ and it is at $u \to \infty$
for $p>4$. Now to minimize the area \eqref{areareduced} we use the equation of
motion which has the form,
\be\label{eom}
\dot u =
\frac{4-p}{2\sqrt{Q_1}}\sqrt{\left(\frac{u_{\ast}}{u}\right)^{\frac{4}{4-p}} 
-1}
\ee  
where $u_{\ast}$ is the constant of motion. The solution of this equation has
the form that $u$ starts from 0 or $\infty$ (for $p<4$ or $p>4$) i.e., at the
boundary and then comes all the way upto $u=u_{\ast}$, where there is a
turning point ($\dot u =0$) and goes back again to $u=0$ or $\infty$. 
The equation 
\eqref{eom} can be easily integrated to obtain the unknown constant of motion
$u_{\ast}$ in terms of the width of the strip $\ell$ as,
\be\label{eomsoln}
u_{\ast} = \frac{4 \ell}{\sqrt{Q_1\pi}} \frac{\Gamma\left(\frac{8-p}{4}\right)}
{\Gamma\left(\frac{6-p}{4}\right)}
\ee
Now substituting \eqref{eom} into \eqref{areareduced} and replacing $u$ by a
dimensionless variable $x=u/u_{\ast}$, we obtain the area as,
\be\label{area1}
{\rm Area}(\gamma_A) = \frac{2Q_1 Q_2^{\half} L^p}{4-p}
u_{\ast}^{\frac{2-p}{4-p}} \int
\frac{dx}{x^{\frac{2}{4-p}}\sqrt{1-x^{\frac{4}{4-p}}}}
\ee
We would like to remark that the above area \eqref{area1} is actually
divergent near the boundary $u=0$ for $p<4$ and $u \to \infty$ for $p>4$. So,
we put a cut-off $u_{\rm min}$, $u_{\rm max}$ for $p<4$ and $p>4$
respectively. So, putting the proper integration limit, the holographic
EE can be obtained from \eqref{EE} as,
\be\label{EE1}
S_A = \frac{1}{4G_N^{p+3}}\frac{2Q_1 Q_2^{\half} L^p}{4-p}
u_{\ast}^{\frac{2-p}{4-p}} \int_{\frac{u_{\rm min}}{u_{\ast}},
\frac{u_{\rm max}}{u_{\ast}}}^1
\frac{dx}{x^{\frac{2}{4-p}}\sqrt{1-x^{\frac{4}{4-p}}}}
\ee
where the lower limits $u_{\rm min}/u_{\ast}$ and $u_{\rm max}/u_{\ast}$ refer
to $p<4$ and $p>4$ cases respectively. The integral in \eqref{EE1} can be
easily evaluated for $p\neq 2$. Actually, for $p \neq 2$, we write the
results in two parts, one away from the boundary which is in general finite
and the other near the boundary when we take 
$u_{\rm min} \to 0$ (for $p<4$) or $u_{\rm max} \to \infty$ (for $p>4$). To
extract the part near the boundary we expand the sqare root in the integrand 
for $x$ nearly zero for $p<4$ and $1/x$ nearly zero for $p>4$. The leading
contribution of the integral near the boundary has the form,
\be\label{intbound}
\int^{\frac{u_{\rm min}}{u_\ast}, \frac{u_{\rm max}}{u_\ast}}   
 \frac{dx}{x^{\frac{2}{4-p}}\sqrt{1-x^{\frac{4}{4-p}}}} \approx 
\frac{4-p}{2-p} \left(\frac{u_{\rm min,\,max}}{u_\ast}\right)^{\frac{2-p}{4-p}}
\ee
The part away from the boundary is the regularized integral and has finite
contribution given by,
\bea\label{awaybound}
& &\int_{0,\infty}^1  
\frac{dx}{x^{\frac{2}{4-p}}\sqrt{1-x^{\frac{4}{4-p}}}} - \frac{4-p}{2-p} 
\left(\frac{u_{\rm min,\,max}}{u_\ast}\right)^{\frac{2-p}{4-p}}\nn
&=& \frac{4-p}{4} \frac{\sqrt{\pi} \Gamma\left(\frac{2-p}{4}\right)}
{\Gamma\left(\frac{4-p}{4}\right)}
\eea
The lower limits $(0,\,\infty)$ in \eqref{awaybound} refer to $p<4$ and $p>4$
cases. Therefore the holographic EE for the strip on the boundary of F-D$p$
solution has the form,
\be\label{HEE}
S_A = \frac{Q_1Q_2^{\half}}{2G_N^{p+3}}L^p\left(\frac{(u_{\rm
      min,\,max})^{\frac{2-p}{4-p}}}{2-p} +
  \frac{\pi^{\frac{1}{4-p}}}{4^{\frac{2}{4-p}}Q_1^{\frac{2-p}{2(4-p)}}}
\frac{\Gamma\left(\frac{2-p}{4}\right)}
{\Gamma\left(\frac{4-p}{4}\right)} \left(\frac{\Gamma\left(\frac{8-p}{4}
\right)}{\Gamma\left(\frac{6-p}{4}\right)}\right)^{\frac{2-p}{4-p}}
\ell^{\frac{2-p}{4-p}} \right)
\ee
In the above $u_{\rm min,\,max}$ refers to $p<4$ and $p>4$ cases. The
holographic EE given in \eqref{HEE} is valid for $p=0,\,1,\,3,\,$ and 5.
As we will discuss, for $p=2$, the holographic EE has a logarithmic 
violation of
the area law \cite{Ogawa:2011bz, Huijse:2011ef}. From \eqref{HEE} we see 
that for $p=0,\,1$, the first term actually vanishes as we take
$u_{\rm min} \to 0$ and $S_A$ is finite for these cases. For $p=0$, $S_A \sim
\ell^{\half}$ and for $p=1$, $S_A \sim L\ell^{\frac{1}{3}}$. In these cases
the EE has the behavior which is in between the logarithmic and 
linear beahaviors.
In fact these two cases are where the hyperscaling violation exponent $\theta
= p - (p-2)/(4-p)$ lies between $d=p+1$ and $d-1 =p$, i.e. $d-1 < \theta < d$
and therefore as noted in \cite{Dong:2012se}, there are new violations of area law
indicating that there are new phases for these kind of systems. On the other
hand for $p=3$, we find $S_A = \alpha_3 L^p/u_{\rm min}^{p-\theta} 
+ \beta_3 L^p/\ell^{p-\theta}$, where $\alpha_3$ and 
$\beta_3$ are known constants (as given in \eqref{HEE}) and 
$\theta = 2$. For $p=5$, we have $S_A = \alpha_5 
L^p/u_{\rm max}^{p-\theta} + \beta_5 L^p/\ell^{p-\theta}$,
where $\alpha_5$, $\beta_5$ are known constants (as given in \eqref{HEE}) 
and $\theta = 8$. Thus for $p=3,5$
we get the usual area law of the holographic EE \cite{Bombelli:1986rw, 
Srednicki:1993im} for a system having Lifshitz
scaling with hyperscaling violation. 

For $p=2$, the relation \eqref{eomsoln} is still valid and from there we get
\be\label{eomsolnp2}
u_{\ast} = \frac{2\ell}{\sqrt{Q_1}}
\ee
However, the integration in \eqref{EE1} gives a logarithmically divergent 
contribution. Using \eqref{eomsolnp2}, we find the holographic EE for $p=2$
case as,
\be\label{EE1p2}
S_A = \frac{Q_1Q_2^{\half}}{4G_N^5} L^2 \log\frac{4\ell}{\sqrt{Q_1}u_{\rm
    min}}
\ee
We thus find that in this case the holographic EE gives a logarithmic
violation of the area law. This was interpreted in the boundary theory as the
presence of hidden Fermi surface \cite{Ogawa:2011bz, Huijse:2011ef}.

Note that the holographic EE \eqref{HEE}, \eqref{EE1p2} have been calculated 
here using the supergravity
configuration \eqref{fdpinnewcoord} which is valid under certain range of
parameter $u$ as mentioned earlier. These ranges have been discussed in 
\cite{Dey:2012tg} and here we discuss them in the context of EE. 

Let us first consider the case of odd $p$. For $p=1$, the gravity
description \eqref{fdpinnewcoord} is valid when $Q_2\ll Q_1$ and 
$u\gg1/Q_2^{3/2}\gg 1/Q_1^{3/2}$ and so the EE \eqref{HEE} is valid also in
this range. However, when $Q_2\geq Q_1$, we have to go to the S-dual frame and
the gravity description is then valid for $u \gg 1/Q_1^{3/2} \gg 1/Q_2^{3/2}$.
We calculated the EE also in the S-dual frame and found that they have exactly
the same form \eqref{HEE} as calculated from the original description except
that the range of validity is different. Also the cut-off parameter $u_{\rm min}$
can be taken to close to the boundary only if we have both
$Q_1,\, Q_2 \gg 1$. For $p=3$, the supergravity description
\eqref{fdpinnewcoord} as well as the holographic EE \eqref{HEE} are valid 
in the
range $1/Q_1^{1/6} \ll u \ll Q_2^{1/2}$. Here $u_{\rm min}$ can be taken to
close to the boundary if we assume $Q_1 \gg 1$. When $u\leq 1/Q_1^{1/6}$, we
have to go to the S-dual description and again, as in the case of $p=1$, 
we find
that the holographic EE has exactly the same form \eqref{HEE} as in the
original description. In the S-dual frame $u_{\rm min}$ can be taken as close
to the boundary as possible without any restriction to the charges. 
Same thing happens for $p=5$ as well. Here the supergravity description
\eqref{fdpinnewcoord} and the holographic EE \eqref{HEE} are valid in the
range $Q_2^{1/6} \ll u \ll (Q_1Q_2)^{1/4}$. In this case $u_{\rm max}$ can be
taken to close to the boundary if $Q_1Q_2 \gg 1$. However for $u \geq
(Q_1Q_2)^{1/4}$, we have to go to the S-dual description. But the holographic
EE has exactly the same form \eqref{HEE} as in the original description. 
In the S-dual description $u_{\rm max}$ can be taken to the close to the 
boundary without any restriction to the charges.

For even $p$, the situation is slightly different. So, for example, 
for $p=0$, the supergravity description \eqref{fdpinnewcoord} as well as the 
holographic EE \eqref{HEE} are valid in the range 
$1/Q_2 \ll u \ll Q_1^{2/3}/Q_2$ and the cut-off parameter $u_{\rm min}$ can be 
taken close 
to the boundary if $Q_2 \gg 1$. However, when $u\geq Q_1^{2/3}/Q_2$, the
dilaton is large and we have to uplift the solution to M-theory. Since here
the boundary theory has one dimension higher, we find that the holographic EE
in this case has exactly the same form \eqref{HEE} except that the overall
factor $L^0/G_N^{3}$ is replaced by $L^{1}/G_N^{4}$. The same is true
for $p=2$ case as well. Here, the supergravity description 
\eqref{fdpinnewcoord} as well as the holographic EE \eqref{EE1p2} are valid in
the range $u\gg Q_2^{1/4}/Q_1^{1/2}$ along with $Q_2\gg 1$. $u_{\rm min}$ in
this case can be taken to close to the boundary if $Q_1\gg 1$ such that
$Q_2^{1/4}/Q_1^{1/2} \ll 1$. However, when $u \leq Q_2^{1/4}/Q_1^{1/2}$ we
have to uplift the solution to M-theory. Again the holographic EE can be
found to have exactly the same form as \eqref{EE1p2} except that the overall 
factor $L^2/G_N^5$ is replaced by $L^3/G_N^6$.  

Before we conclude this section, we here make a comparison of our results on
holographic EE and those obtained in \cite{Dong:2012se}. The metric used in
\cite{Dong:2012se} has a quite generic form and is not obtained from any 
fundamental theory. In particular, it is not known whether the space-time is
in general stable or not. Whereas our metrics are obtained from the near horizon limit
of some known string theory solutions \cite{Dey:2012tg} which preserve at least
a quarter of the space-time susy and are therefore stable. Although the forms
of the holographic EE given in eqs.(4.24) and (4.26) of \cite{Dong:2012se} are
similar to the forms we obtained in this paper in (14) and (16), the details are
quite different. So, for example, the holographic EE in eqs.(4.24) and (4.26) of
\cite{Dong:2012se} depend on the AdS radius $R$ and a cross-over scale $r_F$, whereas,
the holographic EE in this paper depends on two charge parameters $Q_1$ and $Q_2$
(see (14) and (16)) of the F-strings and D$p$-branes respectively. By comparing them 
we find that these parameters are related as $R^d/r_F^{\theta} = Q_1 Q_2^{d/2}$. However,
to understand the precise relations between them, i.e., how $R$ and $r_F$ are separately
related to the charges $Q_1$ and $Q_2$, we need to obtain the near horizon F-D$p$ solution
as a deformation of a relativistic solution. This at present is not known. The boundary theory
in \cite{Dong:2012se} always lives at $r \to 0$ (in their notation), whereas in this
paper the boundary theory lives at $u \to 0$ for $p<4$, but lives at $u \to \infty$ for
$p>4$. Also note that the various holographic properties of the EE observed in \cite{Dong:2012se}
for generic metric are concretely realized in our string theory solutions for various
values of $p$. Thus, for example, for $p=3,\,5$ ($p=5$ case is not included in \cite{Dong:2012se}),
the EE has the usual area law, for $p=2$, the EE has logarithmic violation of the area law
and for $p=0,\,1$, the EE has new area law violations which is in between linear and 
logarithmic behaviors. The string theory realizations discussed in \cite{Dong:2012se}
are for $z=1$, i.e., for the relativistic cases whereas the string theory realizations we 
have discussed are of Lifshitz-like and non-relativistic. Finally, as the solutions in
\cite{Dong:2012se} are not obtained from string theory, there is no restriction as such on
the radial parameter. Whereas the solutions discussed in this paper are valid as long as
the effective string coupling (the dilaton) and the curvature of the metric (in units of
$\alpha'$) remain small. These give restrictions on the radial parameter $u$ as we have discussed.
We find that even in the strongly coupled region the holographic EE have very similar structures
as those of the original solutions.

\section{Holographic EE at finite temperature}

In this section we compute the holographic EE at finite temperature for the
strip embedded on fixed time slice of the boundary of the F-D$p$ system. For
this purpose we start from the non-extremal F-D$p$ bound state solutions and
then take the near horizon limit. In a suitable coordinate the near horizon
metric in the string frame and the dilaton of this solution take the form,
\bea\label{fdpne}
ds^2 &=& Q_2^{\half} u^{\frac{2-p}{4-p}}\left[-\frac{f(u)dt^2}{Q_1Q_2
u^{\frac{4(5-p)}{4-p}}}
+ \frac{\sum_{i=1}^p(dx^i)^2}{Q_2 u^2} + \frac{(dx^{p+1})^2}{Q_1 u^2}
+ \frac{4}{(4-p)^2}\frac{du^2}{f(u) u^2} + d\Omega_{7-p}^2\right]\nn
e^{2\phi} &=& \frac{Q_2^{\frac{3-p}{2}}}{Q_1} u^{\frac{(6-p)(1-p)}{(4-p)}} 
\eea
where $f(u) = 1 - (u/u_h)^{2(6-p)/(4-p)}$ and $u=u_h$ is the radius of the
event horizon. As before compactifying the above metric on $S^{7-p}$ and
writing the resultant metric in Einstein frame we get, 
\be\label{reducedmetricne}
ds_{p+3}^2 = Q_1^{\frac{2}{p+1}} Q_2
u^{2\frac{p(4-p)-(p-2)}{(4-p)(p+1)}}\left[-\frac{f(u) dt^2}{Q_1Q_2
u^{\frac{4(5-p)}{(4-p)}}} + \frac{\sum_{i=1}^p(dx^i)^2}{Q_2u^2} + 
\frac{(dx^{p+1})^2}{Q_1u^2} + \frac{4}{(4-p)^2}\frac{du^2}{f(u)u^2}\right]
\ee
From this black hole metric we can calculate its temperature and it has the
form
\be\label{temp}
T = \frac{6-p}{4\pi (Q_1Q_2)^{\half} u_h^{\frac{2(5-p)}{4-p}}}
\ee
The thermal entropy or Bekenstein-Hawking entropy which is proportional to 
the area of the black hole can, therefore, be written as,
\be\label{bhentropy}
S_T = \frac{1}{4G_N^{p+3}} (Q_1Q_2)^{\half}V
\left(\frac{4\pi (Q_1Q_2)^{\half}}{6-p}\right)^{\frac{1}{5-p}} 
T^{\frac{1}{5-p}}
\ee
Note that the expression \eqref{bhentropy} for the entropy holds good for
$p<5$. For $p=5$, as we see from \eqref{temp}, the temperature does not depend
on $u_h$ and therefore the thermal entropy in that case is independent of $T$
which implies that the specific heat vanishes. The specific heat is positive
only for $p<5$ cases and therefore we will restrict our discussion only for
those cases.

We now calculate the holographic EE at finite temperature, for the same strip
as in the previous section, from the near
horizon non-extremal F-D$p$ configuration \eqref{reducedmetricne}. The area
expression \eqref{areareduced} in this case will be modified as,
\be\label{areareducedfin}
{\rm Area}(\gamma_A) = (Q_1Q_2)^{\frac{1}{2}}L^p \int_{-\ell}^{\ell} dx^{p+1} 
u^{-\frac{2}{4-p}} \sqrt{1+\frac{4Q_1{\dot u}^2}{(4-p)^2 f(u)}}
\ee
The equation of motion \eqref{eom}, would therefore be modified as,
\be\label{eomfin}
\dot u =
\frac{4-p}{2\sqrt{Q_1}} \sqrt{f(u)}\sqrt{\left(\frac{u_{\ast}}{u}
\right)^{\frac{4}{4-p}} -1}
\ee
As before, introducing a dimensionless variable $x=u/u_{\ast}$, the above
equation can be integrated to obtain,
\be\label{eomsoln1}
\ell = \frac{\sqrt{Q_1}}{4-p}u_{\ast} I\left(\frac{u_{\ast}}{u_h}\right)
\ee
where,
\be\label{integral1}
I\left(\frac{u_{\ast}}{u_h}\right) = \int_0^1 dx \frac{x^{\frac{2}{4-p}}}
{\sqrt{1-\left(x\frac{u_{\ast}}{u_h}\right)^{\frac{2(6-p)}{4-p}}}
\sqrt{1-x^{\frac{4}{4-p}}}}
\ee 
Now writing $dx^{p+1} = du/{\dot u}$ in the area expression
\eqref{areareducedfin} and using ${\dot u}$ from \eqref{eomfin}, the area
expression reduces to,
\be\label{areareducedfin1}
{\rm Area}(\gamma_A) = \frac{2Q_1Q_2^{\half}L^p u_{\ast}^{\frac{2-p}{4-p}}}
{4-p} \tilde I\left(\frac{u_{\ast}}{u_h}\right)
\ee
where
\be\label{integral2}
\tilde I\left(\frac{u_{\ast}}{u_h}\right) = 
\int_{\frac{u_{\rm min}}{u_{\ast}}}^1 dx 
\frac{x^{-\frac{2}{4-p}}}
{\sqrt{1-\left(x\frac{u_{\ast}}{u_h}\right)^{\frac{2(6-p)}{4-p}}}
\sqrt{1-x^{\frac{4}{4-p}}}}
\ee
Note that even though the integral \eqref{integral1} is convergent near the
boundary ($u=0$), the integral \eqref{integral2} is divergent there
and therefore we have put a cut-off $u_{\rm min}$ in $\tilde I$. The
finite temperature EE therefore takes the form,
\be\label{HEEfin}
S_A^{\rm finite} = \frac{Q_1Q_2^{\half}L^p u_{\ast}^{\frac{2-p}{4-p}}}
{2G_N^{p+3}(4-p)} \tilde I\left(\frac{u_{\ast}}{u_h}\right)
\ee
Now in order to express the finite temperature EE in terms of the
temperature and $\ell$, as in the case of thermal entropy \eqref{bhentropy},
we have to evaluate $\tilde I$. However, this integral can not be performed
analytically. It can be evaluated only numerically. For that we have to first
numerically integrate $I$ in \eqref{integral1} to obtain $u_{\ast}$ in terms of
$\ell$ and $u_h$ (which in turn is related to the temperature $T$ by the 
relation \eqref{temp}) and then use that to numerically obtain $S_A^{\rm
  finite}$ from \eqref{HEEfin}. Here we discuss the low and the high
temperature behaviors of $S_A^{\rm finite}$ as done in \cite{Dong:2012se}.
 
From \eqref{temp} we find that $T \sim u_h^{-2(5-p)/(4-p)}$ and so for $p<4$, 
as $u_h \to 0$, i.e. as $u_h$ goes to the boundary, $T \to \infty$ and 
as $u_h \to
\infty$, i.e. in the extremal limit, $T \to 0$ as expected. It is, therefore,
clear that in the small temperature regime, $u_{\ast}/u_h \sim \ell 
T^{(4-p)/2(5-p)} \ll 1$ . Now in this approximation the integral
\eqref{integral1} can be evaluated to the leading order to give $u_{\ast}$ in
terms of $\ell$ which has the same form as given in eq.\eqref{eomsoln}. Once we
have this, the integral \eqref{integral2} can be evaluated by expanding 
the factor in the denominator
$[1-\left(x\frac{u_{\ast}}{u_h}\right)^{\frac{2(6-p)}{4-p}}]^{1/2}$ for small 
$u_{\ast}/u_h$ which in turn gives $S_A^{\rm finite}$ in the form,
\bea\label{HEEfinsmall}
S_A^{\rm finite} &\approx & \frac{Q_1Q_2^{\half}}{2G_N^{p+3}}L^p
\left[\frac{(u_{\rm
      min})^{\frac{2-p}{4-p}}}{2-p} +
  \frac{\pi^{\frac{1}{4-p}}}{4^{\frac{2}{4-p}}Q_1^{\frac{2-p}{2(4-p)}}}
\frac{\Gamma\left(\frac{2-p}{4}\right)}
{\Gamma\left(\frac{4-p}{4}\right)} \left(\frac{\Gamma\left(\frac{8-p}{4}
\right)}{\Gamma\left(\frac{6-p}{4}\right)}\right)^{\frac{2-p}{4-p}}
\ell^{\frac{2-p}{4-p}}\right.\nn
&+&\left. \frac{\sqrt \pi}{8}\frac{\Gamma\left(\frac{14-3p}{4}\right)}
{\Gamma\left(\frac{16-3p}{4}\right)}\left(
\frac{4\Gamma\left(\frac{8-p}{4}\right)}
{\sqrt{Q_1\pi}\Gamma\left(\frac{6-p}{4}\right)}\right)^{\frac{14-3p}{4-p}}
\left(\frac{4\pi(Q_1Q_2)^{\half}}{6-p}\right)^{\frac{6-p}{5-p}}
\ell^{\frac{2-p}{4-p}} \left(\ell
  T^{\frac{4-p}{2(5-p)}}\right)^{\frac{2(6-p)}{4-p}}+\cdots\right]\nn
\eea   
The first two terms in \eqref{HEEfinsmall} is precisely the zero temperature
expression of the holographic EE given earlier in \eqref{HEE} and the 
last term is the
finite temperature correction for small temperature. Note that here our result
is valid only for $p<4$ and that is why in the first term we have only $u_{\rm
  min}$. We point out that the finite temperature correction we have 
obtained here has the same 
structure as given in \cite{Dong:2012se}. Also note that for $p=2$, the zero 
temperature form of the holographic
EE has a logarithmic violation of area law and is given in \eqref{EE1p2} 
but the correction term due to small temperature is still given by the third 
term in \eqref{HEEfinsmall} with $p=2$.      

On the other hand when the temperature is large, $u_h \sim u_{\ast}$. In this
case both the integrals $I$ given in \eqref{integral1} and $\tilde I$ given in
\eqref{integral2} are dominated by the pole at $u=1$ and therefore have the
same values, i.e., $I \approx \tilde I$. So, substituting $I$ from
\eqref{eomsoln1} as $I = (4-p)\ell/(\sqrt{Q_1}u_{\ast})$ into the holographic
EE at finite temperature in \eqref{HEEfin} and then expressing $u_h \sim
u_{\ast}$ in terms of temperature from \eqref{temp} we obtain,
\be\label{HEEfinbig}
S_A^{\rm finite} \approx \frac{1}{4G_N^{p+3}} (Q_1Q_2)^{\half} (L^p 2\ell)
\left(\frac{4\pi (Q_1Q_2)^{\half}}{6-p}\right)^{\frac{1}{5-p}} 
T^{\frac{1}{5-p}}
\ee
This has precisely the same form as the Bekenstein-Hawking entropy given in 
\eqref{bhentropy} if we identify $(L^p 2\ell) = V$. We thus find that the
holographic EE at finite temperature indeed makes a cross over to the thermal
entropy as the temperature is increased \cite{Dong:2012se, Swingle:2011mk}.

We remark that the finite temperature extensions of the holographic EE for the
generic hyperscaling violating geometries have also been discussed in \cite{Dong:2012se},
we here give a concrete string theoretic realizations of those from the non-extremal
F-D$p$ solutions. We have given the exact expressions of the low temperature (28)
and the high temperature (29) behavior of the holographic EE, whereas, in \cite{Dong:2012se},
an approximate expressions are given showing the temperature dependence of the holographic 
EE at finite temperature. As before our expressions depend on the charges of the F-strings,
$Q_1$ and D$p$-branes, $Q_2$, but in \cite{Dong:2012se}, the corresponding expressions
depend on the AdS radius $R$ and the cross-over scale $r_F$.

\section{Conclusion}

To conclude, in this paper we have computed the holographic entanglement
entropy of the near horizon geometry of the threshold, 1/4 BPS F-D$p$ 
($p\neq 4$) bound state solutions of type II string theories using the 
prescription of Ryu and Takayanagi \cite{Ryu:2006bv, Ryu:2006ef}. 
The geometry was shown earlier to have
a Lifshitz scaling with hyperscaling violation and the corresponding boundary
theory may describe certain condensed matter systems near quantum critical
point. We have computed the holographic EE for these systems for both zero
and finite temperature. For $p=0,1$ we have found that the holographic EE 
are finite and have new area law violations having behaviors in between the 
linear and logarithmic behaviors. As indicated in \cite{Dong:2012se}, this 
may imply new
phases in the dual theory and we have a string theory realizations of that.
For $p=2$, we have a logarithmic violations of area law and this indicates
that the dual theory describes compressible metallic states with hidden Fermi
surface. For $p=3,5$ we have the usual area law. We have extended our results
for the non-extremal F-D$p$ system and obtained the Hawking temperature and
the thermal or Bekenstein-Hawking entropy of these systems. We found that only
for $p<4$, the specific heat of the system is positive and so we calculated
the holographic EE for F-D$p$ system only for $p<4$ at finite temperature. 
It has been noted that in general the finite temperature holographic EE can 
not be
expressed in a closed analytic form. Some of the integrals in the expression
can only be computed numerically. However, we have discussed the low and the
high temperature behaviors of the Holographic EE expressions at finite
temperature. We found that as the temperature of the system is increased the
Holographic EE makes a cross over to the thermal entropy of the system
indicating the existence of a cross over function \cite{Swingle:2011mk}
interpolating between the entanglement and thermal entropy.

\section*{Acknowledgements}

One of the authors (PD) would like to acknowledge thankfully the financial
support of the Council of Scientific and Industrial Research, India
(SPM-07/489 (0089)/2010-EMR-I).


\vspace{.5cm}

\begin{thebibliography}{99} 
 
%\cite{Dey:2012tg} 
\bibitem{Dey:2012tg}
  P.~Dey and S.~Roy,
  ``Lifshitz-like space-time from intersecting branes in string/M theory,''
  JHEP {\bf 06}, 129 (2012), [arXiv:1203.5381 [hep-th]].
  %%CITATION = ARXIV:1203.5381;%%

%\cite{Bombelli:1986rw}
\bibitem{Bombelli:1986rw} 
  L.~Bombelli, R.~K.~Koul, J.~Lee and R.~D.~Sorkin,
  ``A Quantum Source of Entropy for Black Holes,''
  Phys.\ Rev.\ D {\bf 34}, 373 (1986).
  %%CITATION = PHRVA,D34,373;%%

%\cite{Srednicki:1993im}
\bibitem{Srednicki:1993im} 
  M.~Srednicki,
  ``Entropy and area,''
  Phys.\ Rev.\ Lett.\  {\bf 71}, 666 (1993)
  [hep-th/9303048].
  %%CITATION = HEP-TH/9303048;%%

%\cite{Holzhey:1994we}
\bibitem{Holzhey:1994we} 
  C.~Holzhey, F.~Larsen and F.~Wilczek,
  ``Geometric and renormalized entropy in conformal field theory,''
  Nucl.\ Phys.\ B {\bf 424}, 443 (1994)
  [hep-th/9403108].
  %%CITATION = HEP-TH/9403108;%%

%\cite{Calabrese:2004eu}
\bibitem{Calabrese:2004eu} 
  P.~Calabrese and J.~L.~Cardy,
  ``Entanglement entropy and quantum field theory,''
  J.\ Stat.\ Mech.\  {\bf 0406}, P06002 (2004)
  [hep-th/0405152];
  %%CITATION = HEP-TH/0405152;%%
%\cite{Calabrese:2005zw}
%\bibitem{Calabrese:2005zw} 
 % P.~Calabrese and J.~L.~Cardy,
  ``Entanglement entropy and quantum field theory: A Non-technical 
introduction,''
  Int.\ J.\ Quant.\ Inf.\  {\bf 4}, 429 (2006)
  [quant-ph/0505193];
  %%CITATION = QUANT-PH/0505193;%%
%\cite{Calabrese:2009qy}
%\bibitem{Calabrese:2009qy} 
%  P.~Calabrese and J.~Cardy,
  ``Entanglement entropy and conformal field theory,''
  J.\ Phys.\ A A {\bf 42}, 504005 (2009)
  [arXiv:0905.4013 [cond-mat.stat-mech]].
  %%CITATION = ARXIV:0905.4013;%%

%\cite{Eisert:2008ur}
\bibitem{Eisert:2008ur} 
  J.~Eisert, M.~Cramer and M.~B.~Plenio,
  ``Area laws for the entanglement entropy - a review,''
  Rev.\ Mod.\ Phys.\  {\bf 82}, 277 (2010)
  [arXiv:0808.3773 [quant-ph]].
  %%CITATION = ARXIV:0808.3773;%%

%\cite{Nishioka:2009un}
\bibitem{Nishioka:2009un} 
  T.~Nishioka, S.~Ryu and T.~Takayanagi,
  ``Holographic Entanglement Entropy: An Overview,''
  J.\ Phys.\ A A {\bf 42}, 504008 (2009)
  [arXiv:0905.0932 [hep-th]].
  %%CITATION = ARXIV:0905.0932;%%

%\cite{Takayanagi:2012kg}
\bibitem{Takayanagi:2012kg} 
  T.~Takayanagi,
  ``Entanglement Entropy from a Holographic Viewpoint,''
  Class.\ Quant.\ Grav.\  {\bf 29}, 153001 (2012)
  [arXiv:1204.2450 [gr-qc]].
  %%CITATION = ARXIV:1204.2450;%%

\bibitem{Vidal}
G.~Vidal, J.~I.~Lattore, E.~Rico, and A.~Kitaev, ``Entanglement in quantum
critical phenomena,'' Phys. Rev. Lett. {\bf 90}, 227902 (2003)
[quant-phys/0211074]. 

\bibitem{Peschel}
I.~Peschel, ``On the entanglement entropy of a XY spin chain,''
JSTAT P12005 (2004) [cond-mat/0410416].

\bibitem{Its}
A.~R.~Its, B.~Q.~Jin, V.~E.~Korepin, ``Entanglement in XY spin chain,''
J. Phys. A {\bf 38}, 2975 (2005) [quant-ph/0409027].

%\cite{Kitaev:2005dm}
\bibitem{Kitaev:2005dm} 
  A.~Kitaev and J.~Preskill,
  %``Topological entanglement entropy,''
  Phys.\ Rev.\ Lett.\  {\bf 96}, 110404 (2006)
  [hep-th/0510092].
  %%CITATION = HEP-TH/0510092;%%

\bibitem{Levin}
M.~Levin and X.~G.~Wen, ``Detecting topological order in a ground state wave
function,'' Phys. Rev. Lett. {\bf 96}, 110405 (2006) [cond-mat/0510613].

%\cite{Ryu:2006bv}
\bibitem{Ryu:2006bv} 
  S.~Ryu and T.~Takayanagi,
  ``Holographic derivation of entanglement entropy from AdS/CFT,''
  Phys.\ Rev.\ Lett.\  {\bf 96}, 181602 (2006)
  [hep-th/0603001].
  %%CITATION = HEP-TH/0603001;%%

%\cite{Ryu:2006ef}
\bibitem{Ryu:2006ef} 
  S.~Ryu and T.~Takayanagi,
  ``Aspects of Holographic Entanglement Entropy,''
  JHEP {\bf 0608}, 045 (2006)
  [hep-th/0605073].
  %%CITATION = HEP-TH/0605073;%%

%\cite{Maldacena:1997re}
\bibitem{Maldacena:1997re} 
  J.~M.~Maldacena,
  ``The Large N limit of superconformal field theories and supergravity,''
  Adv.\ Theor.\ Math.\ Phys.\  {\bf 2}, 231 (1998)
  [hep-th/9711200].
  %%CITATION = HEP-TH/9711200;%%

%\cite{Aharony:1999ti}
\bibitem{Aharony:1999ti} 
  O.~Aharony, S.~S.~Gubser, J.~M.~Maldacena, H.~Ooguri and Y.~Oz,
  ``Large N field theories, string theory and gravity,''
  Phys.\ Rept.\  {\bf 323}, 183 (2000)
  [hep-th/9905111].
  %%CITATION = HEP-TH/9905111;%%

%\cite{Klebanov:2007ws}
\bibitem{Klebanov:2007ws} 
  I.~R.~Klebanov, D.~Kutasov and A.~Murugan,
  ``Entanglement as a probe of confinement,''
  Nucl.\ Phys.\ B {\bf 796}, 274 (2008)
  [arXiv:0709.2140 [hep-th]].
  %%CITATION = ARXIV:0709.2140;%%

%\cite{Pakman:2008ui}
\bibitem{Pakman:2008ui} 
  A.~Pakman and A.~Parnachev,
  ``Topological Entanglement Entropy and Holography,''
  JHEP {\bf 0807}, 097 (2008)
  [arXiv:0805.1891 [hep-th]].
  %%CITATION = ARXIV:0805.1891;%%

%\cite{Arean:2008az}
\bibitem{Arean:2008az} 
  D.~Arean, P.~Merlatti, C.~Nunez and A.~V.~Ramallo,
  ``String duals of two-dimensional (4,4) supersymmetric gauge theories,''
  JHEP {\bf 0812}, 054 (2008)
  [arXiv:0810.1053 [hep-th]].
  %%CITATION = ARXIV:0810.1053;%%

%\cite{Ramallo:2008ew}
\bibitem{Ramallo:2008ew} 
  A.~V.~Ramallo, J.~P.~Shock and D.~Zoakos,
  ``Holographic flavor in N=4 gauge theories in 3d from wrapped branes,''
  JHEP {\bf 0902}, 001 (2009)
  [arXiv:0812.1975 [hep-th]].
  %%CITATION = ARXIV:0812.1975;%%

%\cite{Asplund:2011cq}
\bibitem{Asplund:2011cq} 
  C.~T.~Asplund and S.~G.~Avery,
  ``Evolution of Entanglement Entropy in the D1-D5 Brane System,''
  Phys.\ Rev.\ D {\bf 84}, 124053 (2011)
  [arXiv:1108.2510 [hep-th]].
  %%CITATION = ARXIV:1108.2510;%%

%\cite{vanNiekerk:2011yi}
\bibitem{vanNiekerk:2011yi} 
  A.~van Niekerk,
  ``Entanglement Entropy in NonConformal Holographic Theories,''
  arXiv:1108.2294 [hep-th].
  %%CITATION = ARXIV:1108.2294;%%

%\cite{Azeyanagi:2009pr}
\bibitem{Azeyanagi:2009pr} 
  T.~Azeyanagi, W.~Li and T.~Takayanagi,
  ``On String Theory Duals of Lifshitz-like Fixed Points,''
  JHEP {\bf 0906}, 084 (2009)
  [arXiv:0905.0688 [hep-th]].
  %%CITATION = ARXIV:0905.0688;%%

\bibitem{Shaghoulian:2011aa}
E.~Shaghoulian,
``Holographic Entanglement Entropy and Fermi Surfaces,''
JHEP {\bf 1205}, 065 (2012)
[arXiv:1112.2702 [hep-th]].

%\cite{Dong:2012se}
\bibitem{Dong:2012se} 
  X.~Dong, S.~Harrison, S.~Kachru, G.~Torroba and H.~Wang,
  ``Aspects of holography for theories with hyperscaling violation,''
  JHEP {\bf 1206}, 041 (2012)
  [arXiv:1201.1905 [hep-th]].
  %%CITATION = ARXIV:1201.1905;%%
 
%\cite{Kim:2012nb}
\bibitem{Kim:2012nb} 
  B.~S.~Kim,
  ``Schr\'odinger Holography with and without Hyperscaling Violation,''
  JHEP {\bf 1206}, 116 (2012)
  [arXiv:1202.6062 [hep-th]].
  %%CITATION = ARXIV:1202.6062;%%

\bibitem{Sachdev}
S.~Sachdev, ``Quantum Phase Transitions,'' Cambridge University Press, 1999.

%\cite{Sachdev:2012dq}
\bibitem{Sachdev:2012dq}
  S.~Sachdev,
  ``The quantum phases of matter,''
  arXiv:1203.4565 [hep-th].
  %%CITATION = ARXIV:1203.4565;%%a

%\cite{Ogawa:2011bz}
\bibitem{Ogawa:2011bz} 
  N.~Ogawa, T.~Takayanagi and T.~Ugajin,
  ``Holographic Fermi Surfaces and Entanglement Entropy,''
  JHEP {\bf 1201}, 125 (2012)
  [arXiv:1111.1023 [hep-th]].
  %%CITATION = ARXIV:1111.1023;%%

%\cite{Huijse:2011ef}
\bibitem{Huijse:2011ef} 
  L.~Huijse, S.~Sachdev and B.~Swingle,
  ``Hidden Fermi surfaces in compressible states of gauge-gravity duality,''
  Phys.\ Rev.\ B {\bf 85}, 035121 (2012)
  [arXiv:1112.0573 [cond-mat.str-el]].
  %%CITATION = ARXIV:1112.0573;%%

%\cite{Swingle:2011mk}
\bibitem{Swingle:2011mk} 
  B.~Swingle and T.~Senthil,
  ``Universal crossovers between entanglement entropy and thermal entropy,''
  arXiv:1112.1069 [cond-mat.str-el].
  %%CITATION = ARXIV:1112.1069;%%


\end{thebibliography}
\end{document}